# Uplink Performance Analysis of a Drone Cell in a Random Field of Ground Interferers

Mohammad Mahdi Azari[1], Fernando Rosas[2,3], Alessandro Chiumento[1], Amir Ligata[4], Sofie Pollin[1]

[1]Department of Electrical Engineering, KU Leuven, Belgium

[2] Centre of Complexity Science and Department of Mathematics, Imperial College London, UK

[3] Department of Electrical and Electronic Engineering, Imperial College London, UK

[4]Motive Customer Experience Solutions, Applications & Analytics, Nokia, Belgium

Email: mahdi.azari@kuleuven.be

***Abstract*—Aerial base stations are a promising technology to increase the capabilities of the existing communication networks. However, the existing analytical frameworks do not sufficiently characterize the impact of ground interferers on the aerial base stations. In order to address this issue, we model the effect of interference coming from the coexisting ground networks on the aerial link, which could be the uplink of an aerial cell served by a drone base station. By considering a Poisson field of ground interferers, we characterize the aggregate interference experienced by the drone. This result includes the effect of the drone antenna pattern, the height-dependent shadowing, and various types of environment. We show that the benefits that a drone obtains from a better line-of-sight (LoS) at high altitudes is counteracted by a high vulnerability to the interference coming from the ground. However, by deriving the link coverage probability and transmission rate we show that a drone base station is still a promising technology if the overall system is properly dimensioned according to the given density and transmission power of the interferers. Particularly, our results illustrate how the benefits of such network is maximized by defining the optimal drone altitude and signal-to-interference (SIR) requirement.**

## I. Introduction

Aerial communication platforms are currently receiving a lot of attention from the academic and industrial communities as potential solutions for enhancing the performance of ground-based networks. Significant research efforts are being spent in exploring the benefits of aerial communication platforms in fields such as public safety, traffic offloading, disaster relief services and wireless access provision. Google Loon [1] and Facebook Aquila Drone [2], for instance, aim to bring broadband wireless connectivity to remote areas by employing high altitude platforms. The use of low altitude LTE-A aerial base stations for public safety and capacity enhancement is considered in the ABSOLUTE project [3].

Recent reports indicated that the use of drone base stations in low altitudes provides significant performance improvements compared to terrestrial base stations [4]. These advantages are in large part due to the high probability of line-of-sight (LoS) communication between a ground terminal and a drone. Moreover, the introduction of additional degrees of freedom enabling the operation of a drone at optimal altitude logically results in a performance that is at least as good as a base station at a fixed altitude. As a matter of fact, if the drone operates at a high altitude then the propagation distance to ground devices is large, increasing the path loss. However, a higher altitude leads to higher LoS probability, which in conjunction with the increased path loss generates a critical trade-off for drone deployment. In [4], we have shown that this can be formalized as an optimization problem, with a unique optimal altitude for minimum transmission power. Moreover in [4] we have shown that, although the drone base station requires lower transmission power than a terrestrial base station for a fixed coverage region, it can provide higher sum-rate capacity at a wide range of altitudes.

An undesirable consequence of the favorable propagation conditions, which has been overlooked in most of the existing literature, is that the higher LoS probability at high altitudes can also make the drone more vulnerable to the interference coming from neighboring networks [5], [6]. It is plausible that this feature could severely degrade the performance benefits of aerial base stations. Therefore, a critical challenge for the future of this technology is to provide a clear characterization of the effect of interference on the network performance. Please note that the existing model-based literature either ignores interference [7], [8] or focuses only on the downlink drone cell communications [9]. Therefore, to the best of our knowledge, the effect of interference at the drone base station, and its relationship with network parameters such as transmission power and altitude, is still an open question.

In order to address this issue, our approach is to leverage the existing knowledge about interference characterization in ground-to-ground links [10]–[12], and extend this to consider the ground-to-air case for low altitude drone base stations in urban areas. One key difference with the former is that the statistical properties of the received interference are intrinsically linked to the environment-dependent LoS probability and shadowing which vary with the drone alti-tude and elevation angle [13]. Such characterization provides a realistic insight into the height-dependent performance of a drone cell coexisting with ground cells and enables comparison of results for different environments with different models for both interference and link quality.

Concretely, in this paper we consider an uplink communication network between ground terminals and a drone base station. We provide a novel characterization of the aggregate interference seen by the drone utilizing a directional antenna in the presence of Poisson field of ground interferers. Our results show that the drone is heavily vulnerable to the terrestrial interference due to the high probability of LoS, where areas with many tall obstacles experience less interference. Furthermore, we derive the link coverage probability and transmission rate. Then, we study the maximum performance of the network by optimizing the system parameters such as the drone altitude and signal-to-interference (SIR) requirement. Contrasting with previous works, these results show that including the impact of interference leads to a higher performance for denser environments at high altitudes.

The rest of this paper is organized as follows. Section II presents the network model, and then Section III analyzes the aggregate interference. The coverage probability and transmission rate is derived and investigated in Section IV. Our conclusion is finally presented in Section V.

## II. Network Model

We consider an uplink wireless communication system where a drone acting as an aerial base station is located at altitude $h$ as illustrated in Figure 1. A ground terminal communicates with the drone in the presence of a random field of interferers distributed according to a Poisson point process (PPP) of a fixed density $\lambda$. It is to be noted that the density of the interfering terminals $\lambda$ can be affected by various system parameters including the number of available channels, the user association criterion and MAC protocol. The specific scenarios considered could represent an aerial base station that coexists with a terrestrial network, using the same band. All active transmitters on the ground, hence, create interference to the aerial cell.

An arbitrary ground user equipment (UE) is located at a distance $d$ from the drone and $r$ represents the distance between the UE and the projection of the drone on the ground surface O. The UE forms an angle of $\varphi$ with the drone which is the complement of the elevation angle (see Figure 1). We assume that the UE employs an omni-directional antenna for communication, while the drone is equipped with a directional antenna with beamwidth $\varphi_{\mathrm{A}}$ pointing directly downwards. The ground region covered by the main lobe of the drone antenna forms a disc centered at O, being denoted as $\mathcal{C}$. It is important to note that a UE can be either a source of information or interference.

For the channel model between the UE and the drone, as in [13], we consider LoS and non-LoS (NLoS) links separately characterized by their respective probability of occurrence. The received power from LoS and NLoS links are given by [13]

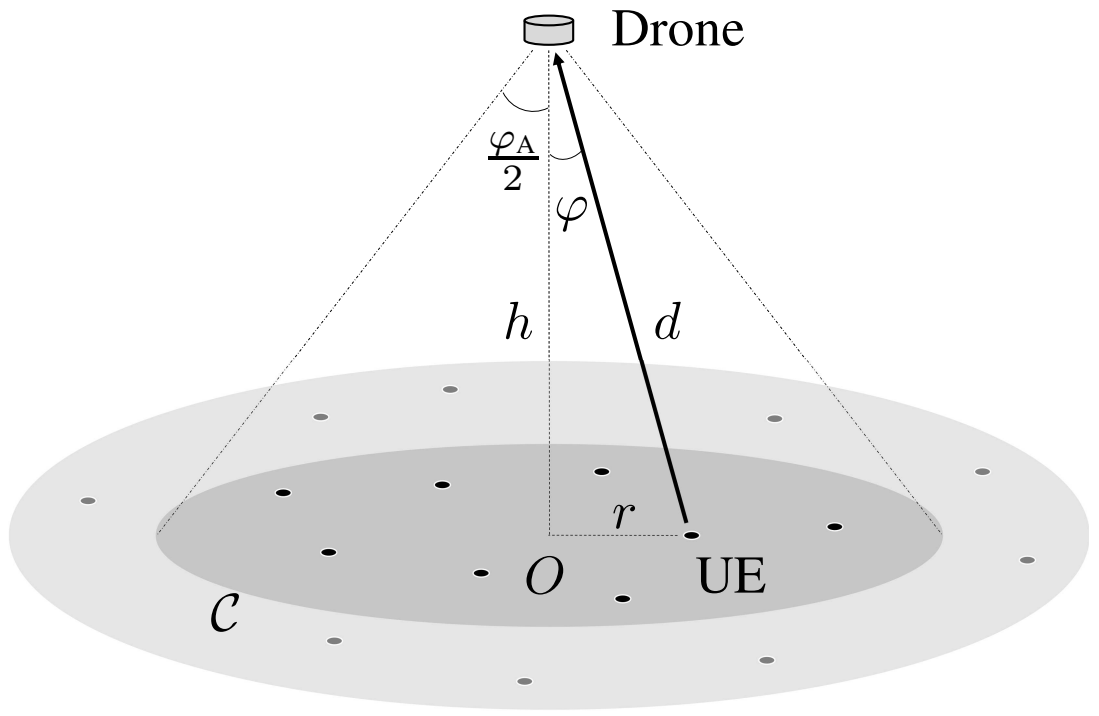

Fig. 1. A typical uplink communication network between a UE and a drone in the presence of random field of ground interferers. The drone utilizes a directional antenna which has the beamwidth of $\varphi_{\mathrm{A}}$.

$$\mathrm{P_r} = \frac{\mathrm{P_t}}{\mathrm{L_f}\Psi_\xi}; \quad \xi \in \{\mathrm{LoS}, \mathrm{NLoS}\}, \quad \mathrm{t} \in \{\mathrm{I}, \mathrm{U}\} \tag{1}$$

where $\mathrm{P_t}$ is the transmit power such that $\mathrm{P_I}$ and $\mathrm{P_U}$ correspond to the source of interference and information respectively, $\mathrm{L_f}$ is the free-space path loss given by

$$\mathrm{L_f} = \left(\frac{4\pi f d}{c}\right)^2 = \mathrm{A_f}\, \frac{h^2}{\cos^2(\varphi)} \;;\; \mathrm{A_f} = \left(\frac{4\pi f}{c}\right)^2, \tag{2}$$

$f$ is the operating frequency and $c$ speed of light. In (1), $\Psi_{\mathrm{LoS}}$ and $\Psi_{\mathrm{NLoS}}$ capture the corresponding excessive path loss and shadowing effect, and follow a log-normal distribution which can be represented as

$$10\log_{10}\Psi_\xi \sim \mathcal{N}(\mu_\xi, \sigma_\xi^2); \quad \xi \in \{\mathrm{LoS}, \mathrm{NLoS}\} \tag{3}$$

In the above equation, $\mu_{\mathrm{LoS}}$ and $\mu_{\mathrm{NLoS}}$ are dependent on $f$ and the type of environment (suburban, urban etc.). Furthermore, $\sigma_{\mathrm{LoS}}$ and $\sigma_{\mathrm{NLoS}}$ are a function of $\varphi$ as

$$\sigma_\xi = a_\xi e^{b_\xi \varphi}; \quad \xi \in \{\mathrm{LoS}, \mathrm{NLoS}\} \tag{4}$$

where $a_{\mathrm{LoS}}$, $b_{\mathrm{LoS}}$, $a_{\mathrm{NLoS}}$, and $b_{\mathrm{NLoS}}$ denote frequency and environment dependent parameters obtained in [13].

The probability of an LoS link between a UE and the drone can be expressed as [13]

$$\mathcal{P}_{\mathrm{LoS}}(\varphi) = \beta_1 \left(\frac{5\pi}{12} - \varphi\right)^{\beta_2}, \tag{5}$$

where $\beta_1$ and $\beta_2$ are frequency and environment dependent parameters and $\varphi$ is in radians. Moreover, the probability of NLoS is equal to $\mathcal{P}_{\mathrm{NLoS}}(\varphi) = 1 - \mathcal{P}_{\mathrm{LoS}}(\varphi)$. We note that from (5), $\mathcal{P}_{\mathrm{LoS}}(\varphi)$ decreases as $\varphi$ increases.

Following, we first investigate and characterize the aggregate interference from ground terminals. Then, we use this result to study the quality of the uplink communication between an arbitrary UE and the drone base station.

## III. Aggregate Interference Model

Here, we statistically characterize the aggregate interference seen by the drone from the ground terminals (i.e. UEs). To this end, we consider $\mathcal{I}$ as a set of interferers within $\mathcal{C}$, which is a random set depending on the point process. $\mathrm{P}_{\mathrm{r},i}$ is the received power from the $i$th intereferer (UE). Hence, the aggregate interference can be written as

$$I_{\mathrm{agg}} = \sum_{i \in \mathcal{I}} \mathrm{P}_{\mathrm{r},i}. \tag{6}$$

Note that in (6), the contribution of interfering terminals located outside of $\mathcal{C}$ is not taken into account due to their small impact. In fact, there are three main reasons why these interferers can be neglected compared to the interferers within $\mathcal{C}$: (I) the probability of LoS, i.e. $\mathcal{P}_{\mathrm{LoS}}(\varphi)$, for the nodes outside of $\mathcal{C}$ is lower, (II) the link length $d$ is larger (resulting in a larger path loss), and (III) the antenna gain is much lower. We hence eliminate the impact of these nodes to simplify the derivations [14]. Moreover, without loss of generality we assume that the gain of the antenna within the main lobe is equal to one to avoid the introduction of another parameter.

The aggregate interference $I_{\mathrm{agg}}$ in (6) is a stochastic process whose distribution depends on the number and location of the interferers as captured by the point process, the statistics of each interfering signal and the signal propagation model. To the best of our knowledge, there is no known closed form expression for the probability distribution function (pdf) of the aggregate interference for the network model introduced in Section II. However, we characterize the mean and variance of $I_{\mathrm{agg}}$ and their dependency on $h$, $\lambda$ and $\varphi_{\mathrm{A}}$ in Section III-A. The results show that the variance compared to the mean is small. For this reason, and for simplicity and tractability, in this paper we employ the mean aggregate interference instead of the actual random value. The network performance simulation using the actual aggregate interference in Section IV-C confirms that the mean aggregate interference is a good approximation of the aggregate interference for the performance analysis.

### A. Aggregate Interference: Mean and Variance

Here, we statistically characterize the aggregate interference from the ground terminals. The mean value of the aggregate interference $\mu_{I_{\mathrm{agg}}} = \mathbb{E}[I_{\mathrm{agg}}]$ is given by the following theorem.

**Theorem 1.** *The mean aggregate interference $\mu_{I_{agg}}$ is given by*

$$\mu_{I_{agg}} = \frac{2\pi\lambda P_I}{A_f} \cdot \Upsilon_\mu(\varphi_A) \tag{7}$$

*where $P_I$ is the transmit power of the interfering nodes,*

$$\Upsilon_\mu(\varphi_A) = \int_0^{\frac{\varphi_A}{2}} \tan(\varphi) \left[ \beta_1 \left(\frac{5\pi}{12} - \varphi\right)^{\beta_2} 10^{\frac{-\mu_{LoS} + v\sigma^2_{LoS}(\varphi)/2}{10}} + \left(1 - \beta_1 \left(\frac{5\pi}{12} - \varphi\right)^{\beta_2}\right) 10^{\frac{-\mu_{NLoS} + v\sigma^2_{NLoS}(\varphi)/2}{10}} \right] d\varphi,$$

*and $v = \frac{\ln(10)}{10}$.*

*Proof.* Assuming that $K$, $\Phi$ and $\Psi$ are random variables denoting, respectively, the number, location and channel statistic of the interferers, one can write

$$\begin{aligned} \mu_{I_{\mathrm{agg}}} &= \mathbb{E}_{K,\Phi,\Psi}\left[\sum_{i=1}^{K} \mathrm{P}_{\mathrm{r},i}\right] \\ &= \mathbb{E}_K \left[ \left( \sum_{i=1}^{k} \mathbb{E}_{\Phi,\Psi}[\mathrm{P}_{\mathrm{r},i}] \right) \Big| K = k \right]. \end{aligned} \tag{8}$$

The received interfering powers $\mathrm{P}_{\mathrm{r},i}$ are identical and assumed to be independent random variables (i.i.d.) with a same distribution denoted as $\mathrm{P}_{\mathrm{r}}$. Therefore, one can rewrite (8) as

$$\begin{aligned} \mu_{I_{\mathrm{agg}}} &= \mathbb{E}_K\left[k \cdot \mathbb{E}_{\Phi,\Psi}[\mathrm{P}_{\mathrm{r}}]\right] \\ &= \mathbb{E}_K[k] \cdot \mathbb{E}_{\Phi,\Psi}[\mathrm{P}_{\mathrm{r}}] = \lambda|\mathcal{C}| \cdot \mathbb{E}_{\Phi,\Psi}[\mathrm{P}_{\mathrm{r}}], \end{aligned} \tag{9}$$

where we take into account the fact that $K$ follows a Poisson distribution with the mean value equal to $\lambda|\mathcal{C}|$ [15]. To evaluate the expectation over $\mathrm{P}_{\mathrm{r}}$ in (9), one can write

$$\begin{aligned} \mathbb{E}_{\Phi,\Psi}[\mathrm{P}_{\mathrm{r}}] &= \int_0^{\frac{\varphi_{\mathrm{A}}}{2}} \mathbb{E}_{\Psi}[\mathrm{P}_{\mathrm{r}}|\Phi = \varphi] \; \mathrm{f}_{\Phi}(\varphi) \; d\varphi \\ &= \frac{2\pi h^2}{|\mathcal{C}|} \int_0^{\frac{\varphi_{\mathrm{A}}}{2}} \frac{\sin(\varphi)}{\cos^3(\varphi)} \mathbb{E}_{\Psi}[\mathrm{P}_{\mathrm{r}}|\Phi = \varphi] \; d\varphi \end{aligned} \tag{10}$$

where $f_{\Phi}(\varphi)$ is the pdf of $\Phi$. We also took into account a uniform distribution of an arbitrary interferer over $\mathcal{C}$ which implies

$$F_{\Phi}(\varphi) \triangleq \mathbb{P}[\Phi \leq \varphi] = \frac{\pi r^2}{|\mathcal{C}|}$$

and hence by using $r = h\tan(\varphi)$ one obtains

$$f_{\Phi}(\varphi) = \frac{\partial}{\partial\varphi} F_{\Phi}(\varphi) = \frac{2\pi h^2}{|\mathcal{C}|} \frac{\sin(\varphi)}{\cos^3(\varphi)}.$$

Now, we notice that

$$\begin{aligned} \mathbb{E}_{\Psi}[\mathrm{P}_{\mathrm{r}}|\Phi = \varphi] &= \mathbb{E}_{\Psi}[\mathrm{P}_{\mathrm{r}}|\Phi = \varphi, \mathrm{LoS}] \cdot \mathcal{P}_{\mathrm{LoS}}(\varphi) \\ &+ \mathbb{E}_{\Psi}[\mathrm{P}_{\mathrm{r}}|\Phi = \varphi, \mathrm{NLoS}] \cdot \mathcal{P}_{\mathrm{NLoS}}(\varphi). \end{aligned} \tag{11}$$

To calculate (11), one finds

$$\mathbb{E}_{\Psi}[\mathrm{P}_{\mathrm{r}}|\Phi = \varphi, \mathrm{LoS}] = \mathbb{E}_{\Psi}\left[\frac{\mathrm{P}_{\mathrm{I}}}{\mathrm{L}_{\mathrm{f}}\Psi_{\mathrm{LoS}}}\right] = \frac{\mathrm{P}_{\mathrm{I}}}{\mathrm{L}_{\mathrm{f}}} \mathbb{E}_{\Psi}\left[\frac{1}{\Psi_{\mathrm{LoS}}}\right]. \tag{12}$$

From (3), we can write

$$\frac{1}{\Psi_{\mathrm{LoS}}} \sim 10^{\frac{\mathcal{N}\left(-\mu_{\mathrm{LoS}}, \sigma^2_{\mathrm{LoS}}\right)}{10}}.$$

Thus

$$\ln\left[\frac{1}{\Psi_{\mathrm{LoS}}}\right] \sim \mathcal{N}\left(-v\mu_{\mathrm{LoS}}, v^2\sigma^2_{\mathrm{LoS}}\right); \quad v = \frac{\ln(10)}{10}$$

which means that $\frac{1}{\Psi_{\mathrm{LoS}}}$ follows a log-normal distribution with parameters $-v\mu_{\mathrm{LoS}}$ and $v^2\sigma^2_{\mathrm{LoS}}$. Therefore, one obtains [16]

$$\mathbb{E}_{\Psi}\left[\frac{1}{\Psi_{\mathrm{LoS}}}\right] = e^{-v\mu_{\mathrm{LoS}} + \frac{v^2\sigma^2_{\mathrm{LoS}}}{2}} = 10^{\frac{-\mu_{\mathrm{LoS}} + v\sigma^2_{\mathrm{LoS}}/2}{10}}. \tag{13}$$

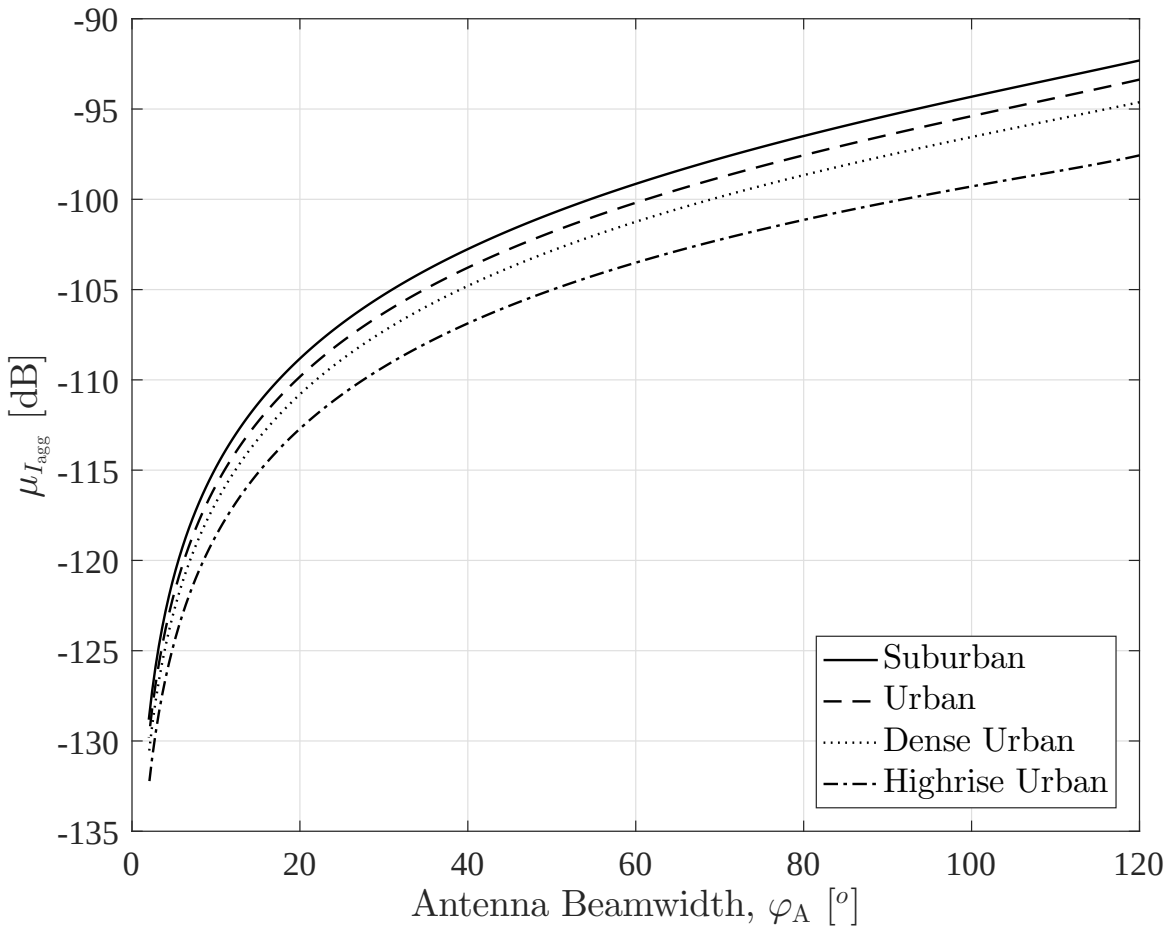

Fig. 2. The mean aggregate interference $\mu_{I_{\text{agg}}}$ increases with drone antenna beamwidth $\varphi_{\text{A}}$, yet the rate of increase slows down for larger antenna beanwidths. Furthermore, the aggregate interference is lower for denser areas.

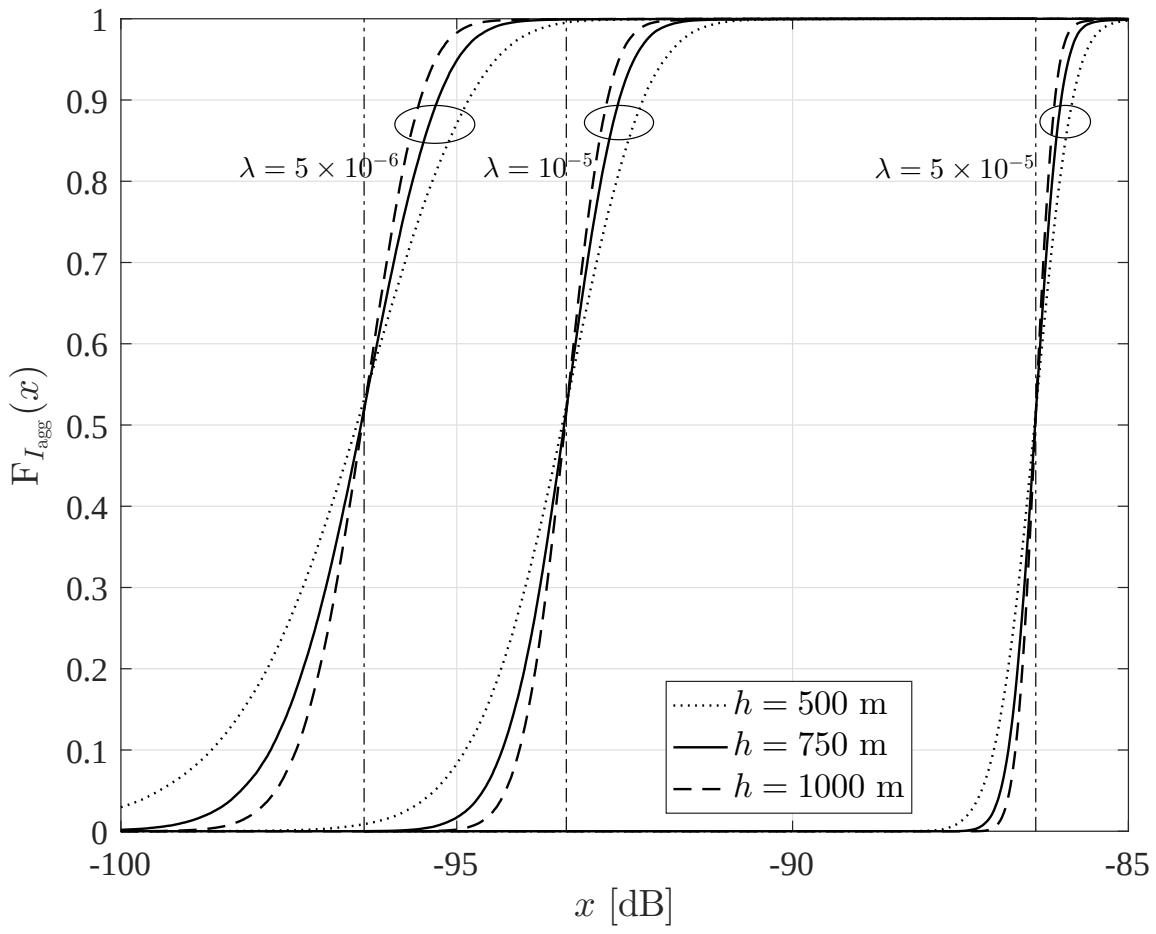

Fig. 3. The aggregate interference $I_{\text{agg}}$ converges to its mean $\mu_{I_{\text{agg}}}$ as $h$ or $\lambda$ increases.

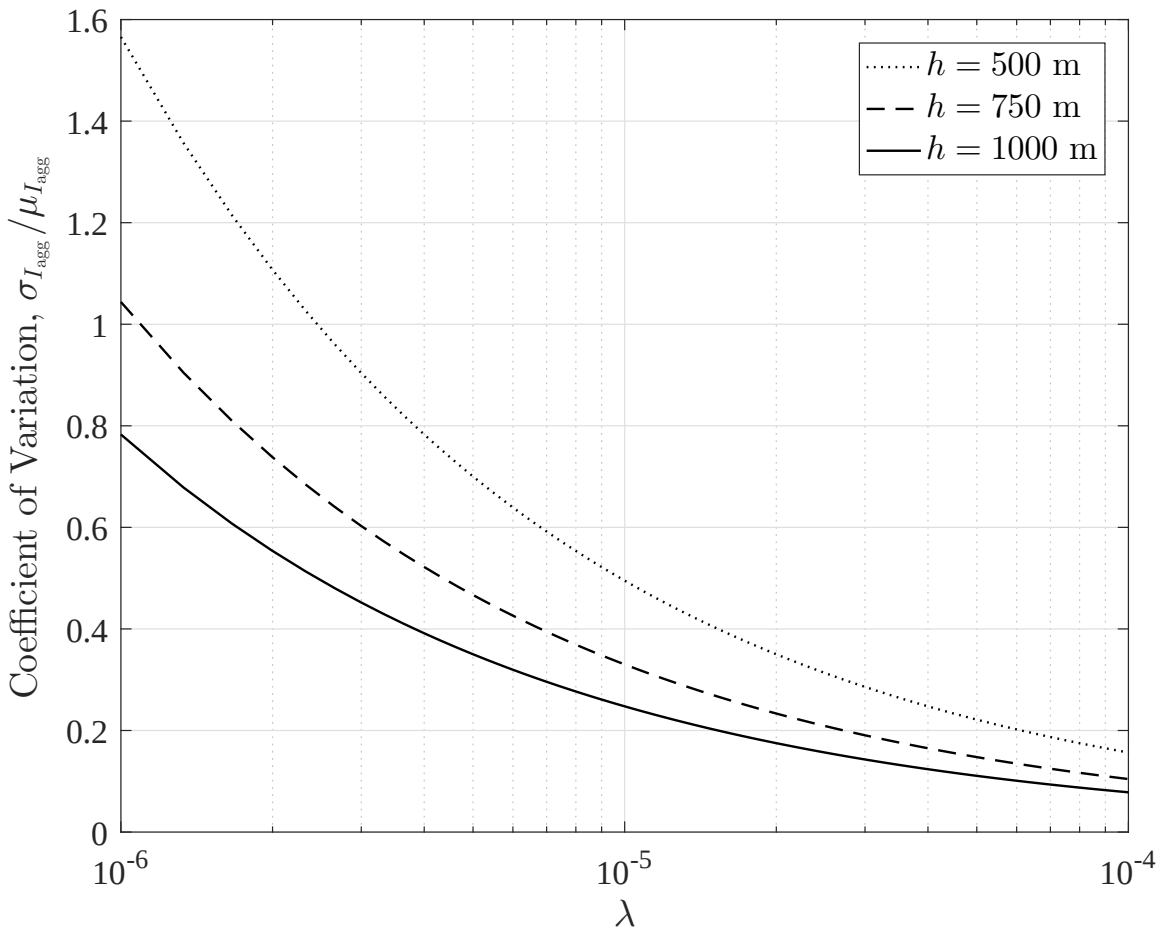

Fig. 4. Larger interferer density $\lambda$ and higher altitude $h$ lead to a lower coefficient of variation $\sigma_{I_{\text{agg}}}/\mu_{I_{\text{agg}}}$.

Similarly, one can write

$$\mathbb{E}_{\Psi}[\text{P}_\text{r}|\Phi = \varphi, \text{NLoS}] = \frac{\text{P}_\text{I}}{\text{L}_\text{f}} \mathbb{E}_{\Psi}\left[\frac{1}{\Psi_{\text{NLoS}}}\right] \tag{14}$$

and

$$\mathbb{E}_{\Psi}\left[\frac{1}{\Psi_{\text{NLoS}}}\right] = 10^{\frac{-\mu_{\text{NLoS}} + v\sigma^2_{\text{NLoS}}/2}{10}}. \tag{15}$$

Therefore, using (2), (5) and (9)–(15), we obtain (7). □

We note that the mean value $\mu_{I_{\text{agg}}}$ in (7) is independent of the drone altitude $h$. In fact, for a given $\varphi_{\text{A}}$ as the drone goes higher, the received power from an interferer decreases proportional to $h^2$ due to path loss. However, at the same time the average number of interferers within $\mathcal{C}$, which is $\lambda|\mathcal{C}|$, grows with the same rate and hence the mean aggregate interference remains constant. Furthermore, we notice that $\mu_{I_{\text{agg}}}$ is greater for a larger antenna beamwidth $\varphi_{\text{A}}$ due to the presence of more interfering nodes within $\mathcal{C}$. In addition, (7) indicates that $\mu_{I_{\text{agg}}}$ is linearly proportional to $\lambda$.

The variance of aggregate interference $\sigma^2_{I_{\text{agg}}}$ is given in the following theorem.

**Theorem 2.** *The variance of the aggregate interference $\sigma^2_{I_{agg}}$ is given by*

$$\sigma^2_{I_{agg}} = \frac{\pi\lambda P_I^2}{A_f^2 h^2} \cdot \Upsilon_\sigma(\varphi_A), \tag{16}$$

*where $P_I$ is the transmit power of the interfering nodes,*

$$\Upsilon_\sigma(\varphi_A) = \int_0^{\frac{\varphi_A}{2}} \sin(2\varphi) \left[\beta_1\left(\frac{5\pi}{12} - \varphi\right)^{\beta_2} 10^{\frac{-\mu_{LoS} + v\sigma^2_{LoS}(\varphi)}{5}} + \left(1 - \beta_1\left(\frac{5\pi}{12} - \varphi\right)^{\beta_2}\right) 10^{\frac{-\mu_{NLoS} + v\sigma^2_{NLoS}(\varphi)}{5}}\right] d\varphi,$$

*and $v = \frac{\ln(10)}{10}$.*

*Proof.* The proof is analogous to the one for Theorem 1 and can be found in Appendix A. □

From Theorems 1 and 2, the coefficient of variation (CV) which is defined as the ratio of $\sigma_{I_{\text{agg}}}$ and $\mu_{I_{\text{agg}}}$, i.e. $\sigma_{I_{\text{agg}}}/\mu_{I_{\text{agg}}}$, is inversely proportional to $\sqrt{\lambda}$ and $h$. This means that as the drone goes higher or the density of the interfering nodes grows, the fluctuation of the aggregate interference around its mean is reduced such that $I_{\text{agg}}$ converges to its mean $\mu_{I_{\text{agg}}}$ for very large $h$ or $\lambda$.

### *B. Numerical Results and Simulation*

In this subsection, we provide the numerical results and simulation by using the following system parameters: $\text{P}_\text{I} = -10$ dB, $\lambda = 10^{-5}$, $h = 500$ m, $\varphi_{\text{A}} = 120^o$, and $f = 2$ GHz, unless otherwise is indicated.

The numerical results in Figure 2 show that the mean interference power increases with the beamwidth $\varphi_{\text{A}}$ due to the presence of more interfering nodes within the main lobe. Note that from the PPP assumption with fixed density $\lambda$, the average number of interferers within $\mathcal{C}$ is equal to $\lambda|\mathcal{C}| = \lambda\pi h^2 \tan^2(\varphi_A/2)$. On the other hand, the interference power seen by the drone above a denser area is lower than in less populated areas because of more blockages and lower LoS probability which deteriorate the received power from the ground terminals. From the figure, the interference growth at higher $\varphi_A$s is slower due to the fact that the interfering signals from the further nodes are much weaker as discussed in the previous subsection.

Moreover, Figure 2 shows that due to a high probability of LoS in a ground-to-air communication link, the power of interference is significantly high which means that a drone is affected strongly by the ground interferers. Although by using directional antenna the cumulative interference can be reduced, the noise power density which is typically less than -150 dB is negligible compared to the interference.

The network simulation in Figure 3 illustrates the cumulative distribution function of the aggregate interference $F_{I_{\text{agg}}}(x)$ for different drone altitudes and ground interferer densities. Each curve is obtained from $10^5$ network realizations. Note that the mean value $\mu_{I_{\text{agg}}}$ at different $h$ is the same while, it is lower for smaller $\lambda$ as derived in Theorem 1. As can be seen from the figure, the relative variation of $I_{\text{agg}}$ around its mean is larger for smaller $\lambda$. To quantify this, we notice that the CV, i.e. $\sigma_{I_{\text{agg}}}/\mu_{I_{\text{agg}}}$, is inversely proportional to $\sqrt{\lambda}$ which is illustrated in Figure 4. Moreover, one can find from this figure that the CV is reduced as $h$ increases such that the CV is inversely proportional to $h$. For this reason, $F_{I_{\text{agg}}}(x)$ in Figure 3 is narrower around its mean for higher altitudes. Generally, the deviation of $I_{\text{agg}}$ from its mean is low and hence using $\mu_{I_{\text{agg}}}$ instead of $I_{\text{agg}}$ is a reasonable approximation.

## IV. Performance Analysis

Here, using the model derived above, we assess the feasibility of having a drone base station coexisting with a terrestrial network, without cell planning. In this case, the drone base station uses the same frequency as the ground cell, and all UEs transmitting to the ground base stations are generating interference. While being a worst case assumption, it is also very realistic, as a cell planning for mobile base stations is not possible. Thus, we have to consider a worst case scenario in which drone base stations fly over a large number of ground cells. Ideally, the drone base stations are capable of using the same frequency as terrestrial networks.

We consider an uplink communication link between a UE and drone base station and evaluate the quality of communication in presence of interfering ground terminals. To this end, we derive the link coverage probability and transmission rate which are adequate choices for characterizing the quality of communication.

### A. Link Coverage Probability

From the derived interference model, it was seen that the interference sensed by the drone is significantly higher than the noise level, and hence we consider an interference-limited channel. Accordingly, the coverage probability of the link between a UE and a drone is defined as

$$\mathcal{P}_{\text{cov}}(\text{T}) = \mathbb{P}[\mathsf{SIR} > \text{T}], \tag{17}$$

where $\mathbb{P}[E]$ is the probability of the event $E$, $\mathsf{SIR}$ is the signal-to-interference ratio and T is an arbitrary $\mathsf{SIR}$ threshold. The coverage probability can be rewritten as

$$\begin{aligned}\mathcal{P}_{\text{cov}}(\text{T}) &= \mathbb{P}[\mathsf{SIR} > \text{T}] \\ &= \mathbb{P}[\mathsf{SIR}_{\text{LoS}} > \text{T}]\mathcal{P}_{\text{LoS}}(\varphi) + \mathbb{P}[\mathsf{SIR}_{\text{NLoS}} > \text{T}]\mathcal{P}_{\text{NLoS}}(\varphi),\end{aligned} \tag{18}$$

where

$$\mathsf{SIR} = \begin{cases} \frac{\text{P}_\text{U}}{\mu_{I_{\text{agg}}}\text{L}_\text{f}\Psi_{\text{LoS}}} & ;\ \text{for LoS} \\ \frac{\text{P}_\text{U}}{\mu_{I_{\text{agg}}}\text{L}_\text{f}\Psi_{\text{NLoS}}} & ;\ \text{for NLoS} \end{cases} \tag{19}$$

In (19), $\text{P}_\text{U}$ is the transmit power of the UE communicating with the drone, and the aggregate interference $I_{\text{agg}}$ is replaced by its mean value $\mu_{I_{\text{agg}}}$ given by (7). Using (5), (18) and (19) we have

$$\begin{aligned}\mathcal{P}_{\text{cov}}(\text{T}) &= \mathbb{P}\left[\Psi_{\text{LoS}} < \frac{\text{P}_\text{U}}{\mu_{I_{\text{agg}}}\text{L}_\text{f}\text{T}}\right]\beta_1\left(\frac{5\pi}{12}-\varphi\right)^{\beta_2} \\ &\quad + \mathbb{P}\left[\Psi_{\text{NLoS}} < \frac{\text{P}_\text{U}}{\mu_{I_{\text{agg}}}\text{L}_\text{f}\text{T}}\right]\left(1-\beta_1\left(\frac{5\pi}{12}-\varphi\right)^{\beta_2}\right) \\ &= Q\left(\frac{\mu_{\text{LoS}}-\psi(\varphi)}{\sigma_{\text{LoS}}(\varphi)}\right)\beta_1\left(\frac{5\pi}{12}-\varphi\right)^{\beta_2} \\ &\quad + Q\left(\frac{\mu_{\text{NLoS}}-\psi(\varphi)}{\sigma_{\text{NLoS}}(\varphi)}\right)\left(1-\beta_1\left(\frac{5\pi}{12}-\varphi\right)^{\beta_2}\right),\end{aligned} \tag{20}$$

where the last equation follows from (3), and

$$\psi(\varphi) = 10\log_{10}\left(\frac{\text{P}_\text{U}}{\mu_{I_{\text{agg}}}\text{L}_\text{f}\text{T}}\right). \tag{21}$$

Note that due to the high interference level, a UE that aims to communicate with a drone base station might need to increase its transmit power $\text{P}_\text{U}$, depending on $\lambda$ and $\varphi_A$, in order to reach a noticeable coverage probability. In (20), larger $\lambda$ and $\varphi_A$ lead to a smaller $\psi(\varphi)$ which deteriorates the coverage probability $\mathcal{P}_{\text{cov}}$. On the other hand, the impact of $h$ on $\varphi$, $\psi(\varphi)$, $\sigma_{\text{LoS}}(\varphi)$, and $\sigma_{\text{NLoS}}(\varphi)$ results in opposite effects on $\mathcal{P}_{\text{cov}}$ such that at an optimum altitude the coverage probability is maximized. This fact is numerically discussed in Subsection IV-C.

### B. Transmission Rate

Following [4], [17] we define the normalized transmission rate over the communication bandwidth as

$$\begin{aligned}\mathcal{R}_\text{n} &\triangleq \log_2(1+\text{T})\cdot\max_h \mathcal{P}_{\text{cov}}(\text{T}) \\ &= \log_2(1+\text{T})\cdot\mathcal{P}_{\text{cov}}^{\max}(\text{T}).\end{aligned} \tag{22}$$

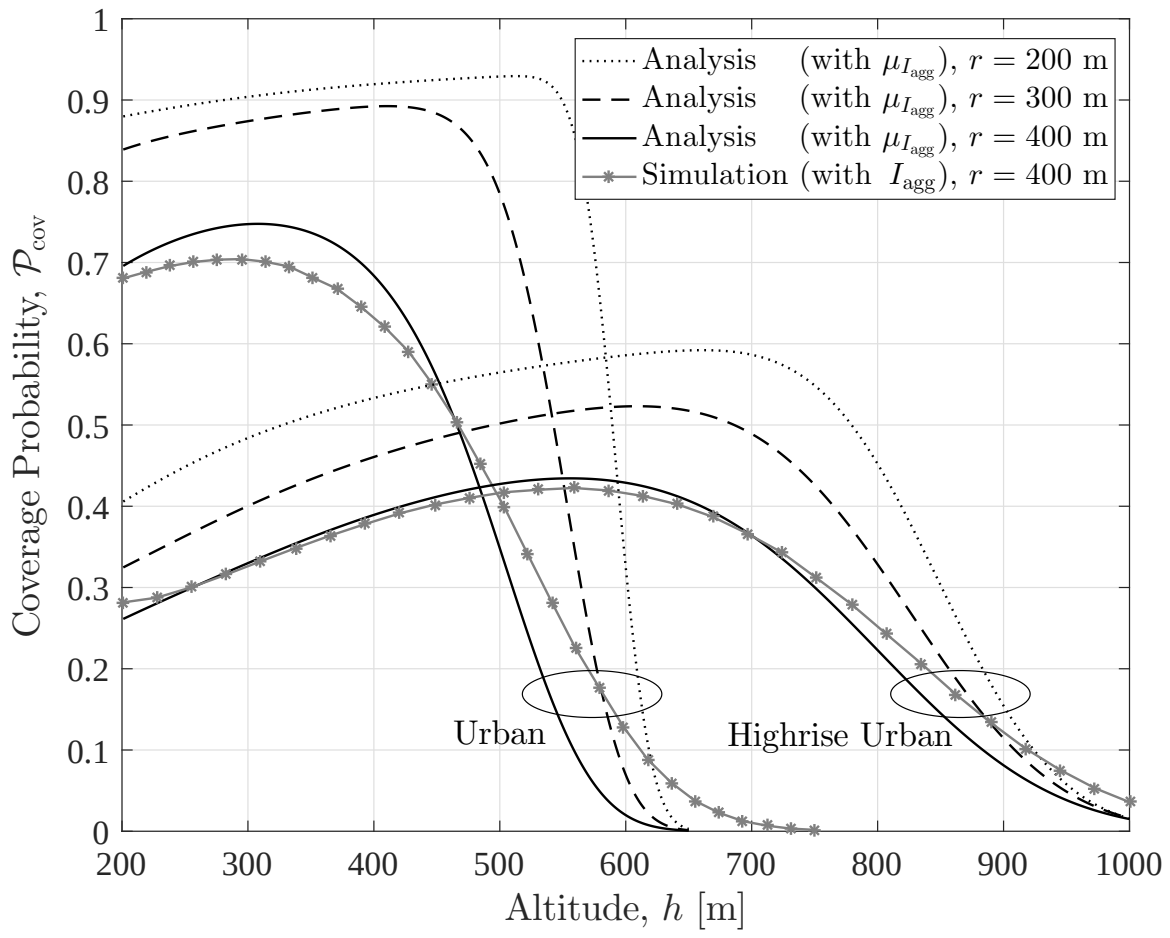

Fig. 5. A denser area shows better performance at high altitudes. The simulation is provided for $r = 400$ m and shows the accuracy of the approximation where we use $\mu_{I_{\mathrm{agg}}}$ instead of $I_{\mathrm{agg}}$ for analyzing the coverage probability $\mathcal{P}_{\mathrm{cov}}$.

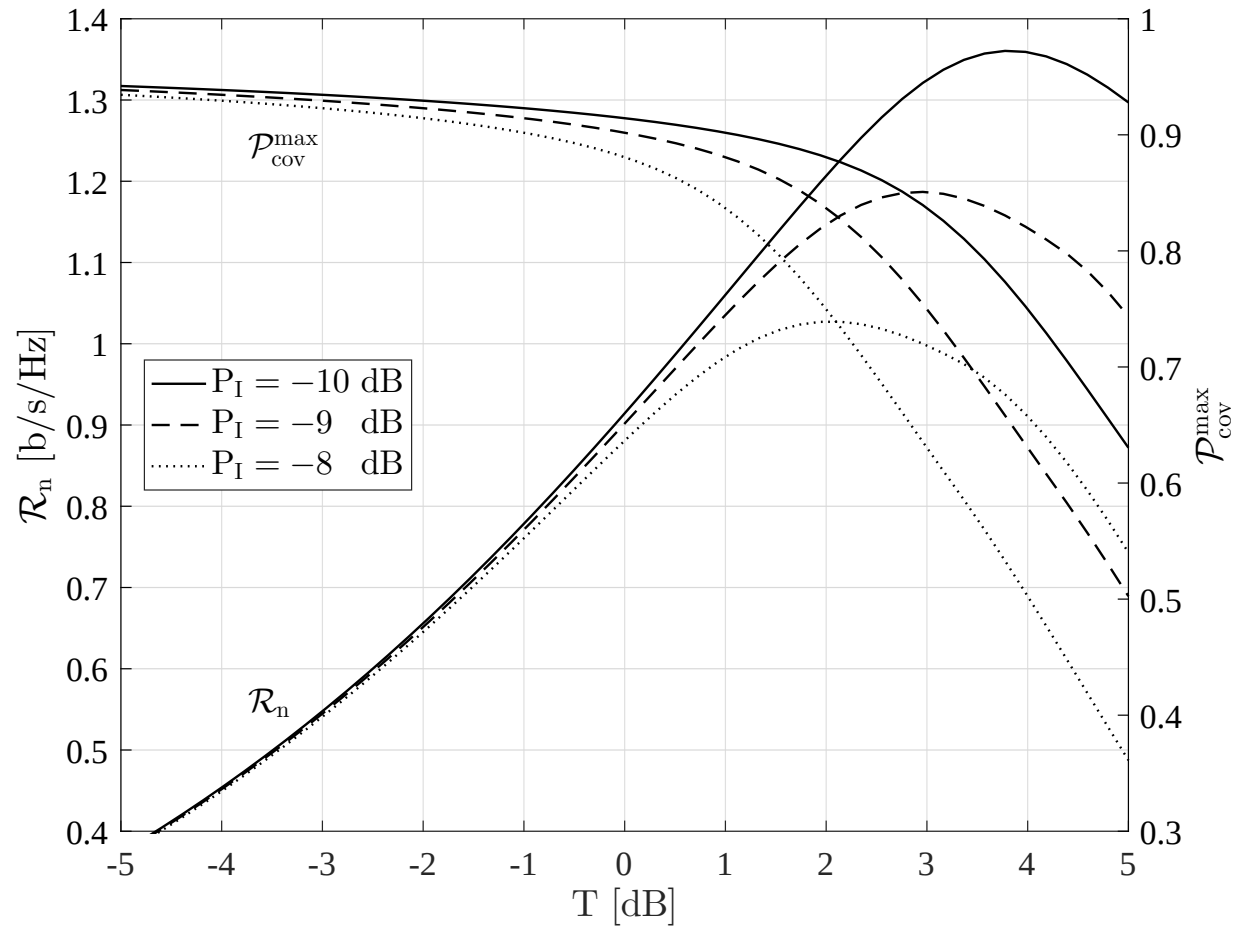

Fig. 6. Transmission rate is maximized at a $\mathcal{P}_{\mathrm{cov}}^{\max}$ lower than its maximum. Lower $\mathrm{P_I}$ results in a higher maximum rate.

Note that (22) assumes that if the UE is in coverage, the normalized data rate over the transmission bandwidth is equal to $\log_2(1 + \mathrm{T})$.

One can find that the first factor in (22), $\log_2(1 + \mathrm{T})$, is directly proportional to T whereas the second factor, $\mathcal{P}_{\mathrm{cov}}^{\max}(\mathrm{T})$, is inversely proportional. This results in an optimal T that maximizes the transmission rate and is evaluated numerically in the next subsection.

### C. Numerical Results and Simulation

In this subsection, we provide the numerical results and simulation for an urban area by using the following system parameters: $\mathrm{P_U} = 0$ dB, $\mathrm{P_I} = -10$ dB, $\lambda = 10^{-5}$, $\varphi_{\mathrm{A}} = 120^o$, $\mathrm{T} = -2$ dB, $r = 200$ m, and $f = 2$ GHz, unless otherwise is indicated.

Figure 5 shows that the coverage probability $\mathcal{P}_{\mathrm{cov}}$ is maximized at an optimum altitude which is dependent on the distance $r$. Indeed, as the drone goes higher the mean aggregate interference power remains the same, while the link length $d$ increases and consequently deteriorates the link SIR. On the other hand, at higher altitudes the drone experiences a higher LoS probability with the UE which leads to a better channel SIR. These two effects are balanced at an optimum altitude where $\mathcal{P}_{\mathrm{cov}}$ reaches its maximum. Shorter $r$ results in a higher optimal altitude and a better coverage probability. As can be seen from the figure, the coverage probability at high altitudes is higher for the denser areas which is due to the lower interference level. Indeed, at low altitudes the link between the UE and drone in Urban area benefits from higher probability of LoS compared to Highrise Urban, however at high altitudes both environments have very high $\mathcal{P}_{\mathrm{LoS}}$ near to one and hence the lower interference level leads to higher coverage probability for Highrise Urban environment. Moreover, the simulation results obtained by $10^5$ network realizations show that the approximation of using $\mu_{I_{\mathrm{agg}}}$ instead of $I_{\mathrm{agg}}$ is a reasonable approach for tractability and simplicity of the network performance analysis.

Figure 6 illustrates the existence of the optimal SIR threshold, i.e. T, at which the link rate is the highest. A higher coverage probability does not necessarily lead to the maximum rate. In fact, the coverage probability $\mathcal{P}_{\mathrm{cov}}(\mathrm{T})$ decreases with T, while $\mathcal{R}_{\mathrm{n}}$ first goes up due to the profound influence of $\log_2(1 + \mathrm{T})$ in (22) and then start decreasing since the reduction in $\mathcal{P}_{\mathrm{cov}}(\mathrm{T})$ becomes dominant. Furthermore, as $\mathrm{P_I}$ increases the maximum of $\mathcal{R}_{\mathrm{n}}$ decreases, however for the low values of T, $\mathcal{R}_{\mathrm{n}}$ remains roughly constant.

## V. Conclusion

In this paper, we considered the uplink performance of a drone cell in the presence of a ground Poisson field of interferers representing multiple co-channel terrestrial cells. We statistically characterized the aggregate interference by its mean value and variance, and showed that the aggregate interference can be well approximated by its mean value for analyzing the network performance without the loss of key features. Our results also indicate that although the aggregate interference is a dominant component and hence limiting the achievable performance, an adequate system dimensioning such as the drone antenna beamwidth, its altitude, the SIR requirement and the transmission power, lead to a considerable network efficiency dependent also on the type of environment. Moreover, we showed that a drone deployed over a dense urban area, has a better performance at high altitudes than a drone deployed in rural areas, due to the lower level of interference and roughly the same probability of LoS.

## APPENDIX
## PROOF OF THEOREM 2

To obtain the variance of $I_{\text{agg}}$ one can write

$$\sigma^2_{I_{\text{agg}}} = \mathbb{E}[I^2_{\text{agg}}] - \left(\mathbb{E}[I_{\text{agg}}]\right)^2. \tag{23}$$

Following the same reasoning as in (8)–(9) we obtain

$$\begin{aligned}
\mathbb{E}[I^2_{\text{agg}}] &= \mathbb{E}_K\left[\mathbb{E}_{\Phi,\Psi}\left(\sum_{i=1}^{k} \mathrm{P}_{\mathrm{r},i}\right)^2 \Big| K = k\right] \\
&= \mathbb{E}_K\left[k\mathbb{E}_{\Phi,\Psi}[\mathrm{P}^2_{\mathrm{r}}] + k(k-1)\left(\mathbb{E}_{\Phi,\Psi}[\mathrm{P}_{\mathrm{r}}]\right)^2\right] \quad (24a) \\
&= \mathbb{E}_K[k]\mathbb{E}_{\Phi,\Psi}[\mathrm{P}^2_{\mathrm{r}}] + \mathbb{E}_K[k(k-1)](\mathbb{E}_{\Phi,\Psi}[\mathrm{P}_{\mathrm{r}}])^2 \\
&= \lambda|\mathcal{C}| \cdot \mathbb{E}_{\Phi,\Psi}[\mathrm{P}^2_{\mathrm{r}}] + (\lambda|\mathcal{C}| \cdot \mathbb{E}_{\Phi,\Psi}[\mathrm{P}_{\mathrm{r}}])^2 \quad (24b) \\
&= \lambda|\mathcal{C}| \cdot \mathbb{E}_{\Phi,\Psi}[\mathrm{P}^2_{\mathrm{r}}] + (\mathbb{E}[I_{\text{agg}}])^2. \quad (24c)
\end{aligned}$$

In (24a) we used the fact that $\mathrm{P}_{\mathrm{r},i}$ are i.i.d, (24b) is obtained by replacing the mean and variance of $K$ with $\lambda|\mathcal{C}|$. Moreover in (24c) the relation in (9) is used.

Using (23) and (24c) yields

$$\sigma^2_{I_{\text{agg}}} = \lambda|\mathcal{C}| \cdot \mathbb{E}_{\Phi,\Psi}[\mathrm{P}^2_{\mathrm{r}}]. \tag{25}$$

Similar to (10)–(12), we can respectively write

$$\begin{aligned}
&\mathbb{E}_{\Phi,\Psi}[\mathrm{P}^2_{\mathrm{r}}] = \frac{2\pi h^2}{|\mathcal{C}|}\int_0^{\frac{\varphi_{\mathrm{A}}}{2}} \frac{\sin(\varphi)}{\cos^3(\varphi)}\mathbb{E}_{\Psi}[\mathrm{P}^2_{\mathrm{r}}|\Phi=\varphi]\, d\varphi, \\
&\mathbb{E}_{\Psi}[\mathrm{P}^2_{\mathrm{r}}|\Phi=\varphi] = \mathbb{E}_{\Psi}[\mathrm{P}^2_{\mathrm{r}}|\Phi=\varphi, \mathrm{LoS}] \cdot \mathcal{P}_{\mathrm{LoS}}(\varphi) \\
&+ \mathbb{E}_{\Psi}[\mathrm{P}^2_{\mathrm{r}}|\Phi=\varphi, \mathrm{NLoS}] \cdot \mathcal{P}_{\mathrm{NLoS}}(\varphi), \\
&\mathbb{E}_{\Psi}[\mathrm{P}^2_{\mathrm{r}}|\Phi=\varphi, \mathrm{LoS}] = \left(\frac{\mathrm{P_I}}{\mathrm{L_f}}\right)^2 \mathbb{E}_{\Psi}\left[\left(\frac{1}{\Psi_{\mathrm{LoS}}}\right)^2\right], \\
&\mathbb{E}_{\Psi}[\mathrm{P}^2_{\mathrm{r}}|\Phi=\varphi, \mathrm{NLoS}] = \left(\frac{\mathrm{P_I}}{\mathrm{L_f}}\right)^2 \mathbb{E}_{\Psi}\left[\left(\frac{1}{\Psi_{\mathrm{NLoS}}}\right)^2\right]. \quad (26)
\end{aligned}$$

Since $\frac{1}{\Psi_{\mathrm{LoS}}}$ follows a log-normal distribution with the parameters $-v\mu_{\mathrm{LoS}}$ and $v^2\sigma^2_{\mathrm{LoS}}$, one can find [16]

$$\mathbb{E}_{\Psi}\left[\left(\frac{1}{\Psi_{\mathrm{LoS}}}\right)^2\right] = 10^{\frac{-\mu_{\mathrm{LoS}} + v\sigma^2_{\mathrm{LoS}}}{5}} \tag{27}$$

and similarly

$$\mathbb{E}_{\Psi}\left[\left(\frac{1}{\Psi_{\mathrm{NLoS}}}\right)^2\right] = 10^{\frac{-\mu_{\mathrm{NLoS}} + v\sigma^2_{\mathrm{NLoS}}}{5}}. \tag{28}$$

Finally, using (2), (5), (25)–(28) and the fact that

$$\sin(2\varphi) = 2\sin(\varphi)\cos(\varphi)$$

we obtain (16).

## REFERENCES

[1] S. Katikala, "Google project loon," *InSight: Rivier Academic Journal*, vol. 10, no. 2, 2014.

[2] Facebook, *Connecting the World from the Sky*. Facebook, Technical Report, 2014.

[3] Absolute (aerial base stations with opportunistic links for unexpected and temporary events). [Online]. Available: http://www.absolute-project.eu

[4] M. M. Azari, F. Rosas, K.-C. Chen, and S. Pollin, "Joint sum-rate and power gain analysis of an aerial base station," in *Globecom Workshops (GC Wkshps), 2016 IEEE*. IEEE, 2016, pp. 1–6.

[5] M. M. Azari, F. Rosas, A. Chiumento, and S. Pollin, "Coexistence of terrestrial and aerial users in cellular networks," in *Globecom Workshops (GC Wkshps)*. to be appeared, 2017. [Online]. Available: https://arxiv.org/abs/1710.03103

[6] M. M. Azari, F. Rosas, and S. Pollin, "Reshaping cellular networks for the sky: The major factors and feasibility," *arXiv preprint arXiv:1710.11404*, 2017.

[7] A. Al-Hourani, S. Kandeepan, and S. Lardner, "Optimal lap altitude for maximum coverage," *Wireless Communications Letters, IEEE*, vol. 3, no. 6, pp. 569–572, 2014.

[8] M. Mozaffari, W. Saad, M. Bennis, and M. Debbah, "Drone small cells in the clouds: Design, deployment and performance analysis," in *Global Communications Conference (GLOBECOM), 2015 IEEE*. IEEE, 2015, pp. 1–6.

[9] M. M. Azari, Y. Murillo, O. Amin, F. Rosas, M.-S. Alouini, and S. Pollin, "Coverage maximization for a poisson field of drone cells," in *IEEE PIMRC*. to be appeared, Oct. 2017. [Online]. Available: https://arxiv.org/abs/1708.06598

[10] H. ElSawy, E. Hossain, and M. Haenggi, "Stochastic geometry for modeling, analysis, and design of multi-tier and cognitive cellular wireless networks: A survey," *IEEE Communications Surveys & Tutorials*, vol. 15, no. 3, pp. 996–1019, 2013.

[11] M. Haenggi, R. K. Ganti *et al.*, "Interference in large wireless networks," *Foundations and Trends® in Networking*, vol. 3, no. 2, pp. 127–248, 2009.

[12] H. ElSawy, A. Sultan-Salem, M.-S. Alouini, and M. Z. Win, "Modeling and analysis of cellular networks using stochastic geometry: A tutorial," *IEEE Communications Surveys & Tutorials*, 2016.

[13] A. Al-Hourani, S. Kandeepan, and A. Jamalipour, "Modeling air-to-ground path loss for low altitude platforms in urban environments," in *Global Communications Conference (GLOBECOM)*. IEEE, 2014, pp. 2898–2904.

[14] M. Rahman and H. Yanikomeroglu, "Enhancing cell-edge performance: a downlink dynamic interference avoidance scheme with inter-cell coordination," *IEEE Transactions on Wireless Communications*, vol. 9, no. 4, pp. 1414–1425, April 2010.

[15] M. M. Azari, A. M. Rabiei, and A. Behnad, "Probabilistic relay assignment strategy for cooperation networks with random relays," *IET Communications*, vol. 8, no. 6, pp. 930–937, 2014.

[16] C. Forbes, M. Evans, N. Hastings, and B. Peacock, *Statistical distributions*. John Wiley & Sons, 2011.

[17] S. Shalmashi, E. Björnson, M. Kountouris, K. W. Sung, and M. Debbah, "Energy efficiency and sum rate tradeoffs for massive mimo systems with underlaid device-to-device communications," *EURASIP Journal on Wireless Communications and Networking*, vol. 2016, no. 1, 2016.